\documentclass[epj]{svjour}
\usepackage{graphics}
\usepackage{color}
\usepackage{ulem}
\usepackage{graphicx}
\usepackage{stfloats}
\usepackage{amsfonts}
\usepackage{threeparttable}
\begin{document}
\title{The $\Delta I$=2 bands in $^{109}$In: possible antimagnetic rotation}
\author{M Wang\inst{1} \and W J Sun\inst{2} \and B H Sun\inst{1,3,}\thanks{\emph{e-mail:} bhsun@buaa.edu.cn} \and J Li\inst{2,}\thanks{\emph{e-mail:} jianli@jlu.edu.cn} \and L H Zhu\inst{1,3,}\thanks{\emph{e-mail:} zhulh@buaa.edu.cn} \and Y Zheng\inst{4} \and G L Zhang\inst{1,3} \and L C He\inst{1} \and W W Qu\inst{1,5} \and F Wang\inst{1} \\
 T F Wang\inst{1,3} \and C Xiong\inst{1} \and C Y He\inst{4} \and G S Li\inst{4} \and J L Wang\inst{4} \and X G Wu\inst{4} \and S H Yao\inst{4} \and C B Li\inst{2} \and H W Li\inst{2} \and S P Hu\inst{6} \and J J Liu\inst{6}
%
}                     
\institute{
School of Physics, Beihang University, Beijing 100191, China\and
College of Physics, Jilin University, Changchun 130012, China\and
Beijing Advanced Innovation Center for Big Data-based Precision Medicine, Beihang University, Beijing, 100083, China\and
China Institute of Atomic Energy, Beijing 102413, China\and
 State Key Laboratory of Radiation Medicine and Protection, School of Radiation Medicine and Protection, Soochow University, Suzhou 215123, China\and College of Physics and Energy, Shenzhen University, Shenzhen 518060, China}
\date{Received: date / Revised version: date}
%
\abstract{
The high-spin structure of $^{109}$In was investigated with the $^{100}$Mo($^{14}$N, 5$n$)$^{109}$In fusion-evaporation reaction at CIAE, Beijing.
Eleven new $\gamma$-rays of $^{109}$In were identified, by which the bandheads of the $\Delta I$=2 rotational bands were confirmed.
The configurations were assigned with the help of the systematic discussion.
Furthermore, the rotational bands are compared with the tilted-axis cranking calculations based on a relativistic mean-field approach.
The rotational bands involving the $1p1h$ excitation to the $\pi$$d_{5/2}$ and $\pi$$g_{7/2}$ orbitals are suggested as  candidates for antimagnetic rotation based on the theoretical results.
} 
\maketitle
\section{Introduction}  \label{intro}
The rotational bands consisting of electric quadrupole transitions are usually related to the rotation of the deformed nucleus around an axis perpendicular to the symmetry axis of the deformed density distribution.
With the development of theoretical and experimental research~\cite{Frauendorfcon,CLARK1992247,197Pb,BALDSIEFEN1992252}, a novel rotation has been found in weakly deformed or near-spherical nuclei, and is interpreted as a result of shears mechanism, i.e., the gradual closing of the angular momentum vector of relatively few high-$j$ proton particles ($j_\pi$) and high-$j$ neutron holes ($j_\nu$).
Such kind of rotation is introduced as magnetic rotation.
Up to now, numerous magnetic rotational bands have been observed in $A \sim 110$ mass region using the HI-13 tandem accelerator at the China Institute of Atomic Energy (CIAE), such as $^{106}$Ag~\cite{M1_106Ag,M1_106Ag_cpc,Shapecoexistence106Ag}, $^{107}$Ag~\cite{107Ag,M1_107Ag}, $^{112}$In~\cite{MR_112In_LXQ,MR112In,MR_112In}, $^{113}$In~\cite{113In,113In_cpl} and $^{115}$In~\cite{MR_115In}.

Antimagnetic rotation (AMR) is another exotic rotation observed in near-spherical nuclei~\cite{Frauendorf2001,meng2013progress,6}.
The angular momentum is increased by simultaneous closing of the two blades of protons and neutrons toward the total angular momentum vector, which is so called ``two-shears-like mechanism".
Because the transverse magnetic moments of the valence nucleons are anti-aligned, there are no $M1$ transitions in antimagnetic rotational bands.
AMR is characterized by weak $E2$ transitions and decreasing $B(E2)$ values with increasing spin, which reflects the nearly spherical core.
The large $\mathfrak {J}^{(2)}/B(E2)$ ratio of the order of 100 $\hbar ^{2}$MeV$^{-1}$$e$$^{-2}$b$^{-2}$, compared with around 10 $\hbar ^{2}$MeV$^{-1}$$e$$^{-2}$b$^{-2}$ for well-deformed nucleus, is also a typical feature~\cite{Frauendorf2001,meng2013progress,6}.

The antimagnetic rotation is expected to be realized in the same mass region with magnetic rotation.
Experimentally, they have been observed simultaneously in the $A \sim 110$ mass region.
Especially for Cd isotopes, the positive parity yrast bands after the alignment of neutrons at sufficiently high frequencies are perfect candidates for the two-shears-like mechanism.
Up to now, the antimagnetic rotational bands have been identified in $^{105}$Cd~\cite{anti105Cd}, $^{106}$Cd~\cite{anti106Cd}, $^{107}$Cd~\cite{anti107Cd}, $^{108}$Cd~\cite{anti106108Cd,antimr_108Cd} and $^{110}$Cd~\cite{anti110Cd}.
For In isotopes, when an additional proton occupies the $g_{7/2}$ or $d_{5/2}$ orbital, the ``two-shears-like mechanism" can also be expected.
In fact, the rotational bands in $^{108,110}$In~\cite{anti108110In,Sun2016}, $^{112}$In~\cite{antiMR_112In} and $^{113}$In~\cite{AMR113In} have been taken as candidates for antimagnetic rotation.

In our previous work~\cite{meng2018}, the triaxial deformation, shape evolution and possible chirality for the dipole bands in $^{109}$In were discussed in detail.
However, it was unclear for the underlying nuclear structure of the $\Delta I$=2 bands.
In this paper,  the level scheme of those bands has been extended by eleven $\gamma$ rays.
The $\Delta I$=2 rotational bands in $^{109}$In are investigated based on the systematic discussion, and the configurations have been suggested.
The experimental results are compared with the tilted axis cranking relativistic mean-field (TAC-RMF) approach~\cite{meng2013progress,6}.
Candidates for antimagnetic rotational bands in $^{109}$In will be discussed.

\begin{figure}
\resizebox{0.5\textwidth}{!}{
\includegraphics[scale=0.65]{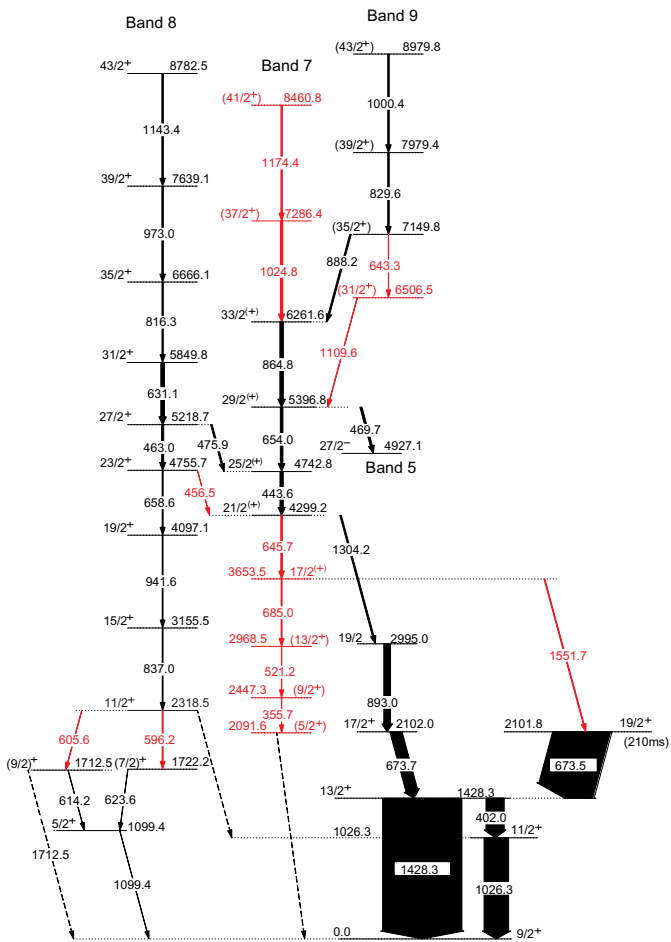}
}
\caption{(color online) Partial level scheme of $^{109}$In proposed in the present work. New transitions and levels are marked as red.}
\label{band78}
\end{figure}

\section{EXPERIMENT AND RESULTS}\label{exp}

The experiment was carried out using the HI-13 tandem accelerator at the China Institute of Atomic Energy (CIAE) in Beijing.
Excited states in $^{109}$In were populated using the $^{100}$Mo($^{14}$N, 5$n$)$^{109}$In fusion-evaporation reaction, and the beam energy was 78 MeV.
The target consisted of a 0.5 mg/cm$^{2}$ foil of $^{100}$Mo with a backing of 10 mg/cm$^{2}$-thick $^{197}$Au.
The $\gamma$ rays were detected by an array composed of nine BGO-Compton-suppressed HPGe detectors, two low-energy photon (LEP) HPGe detectors, and one clover detector.
A total of 84$\times$10$^{6}$ $\gamma$-$\gamma$ coincidence events were sorted into a fully symmetrized $E_{\gamma}$-$E_{\gamma}$ matrix, and analyzed using the software package RADWARE~\cite{Radford} for the $\gamma$-ray coincidence relationship.

The data from the detectors at around 40$^{\circ}$ on one axis and at around 140$^{\circ}$  on the other axis were sorted into an asymmetric DCO matrix.
By analyzed this  asymmetric DCO matrix, the ratios of directional correlation of oriented states (DCO) can be obtained.
The DCO ratios of the known $\gamma$-rays of nuclei produced in the present experiment were taken as the expected value.
When the gate is set on a quadrupole transition, the expected value of stretched quadrupole transitions and pure dipole transitions are around 1.0 and 0.5 in the present array geometry.
Analogously, when the gate is set on dipole transitions, the DCO ratios distribute from 1.5 to 2 for quadrupole transitions and
from 0.5 to 1.3 for dipole transitions.
When the gate is set on pure dipole transitions, the ratios are around 1 for pure dipole transitions.

The partial level scheme focused on the $\Delta I$=2 bands in $^{109}$In is shown in Fig.~\ref{band78}.
The placements of $\gamma$ rays in the level scheme were determined through the observed coincidence relationships, intensity balances, and energy summations.
Compared with the results reported in Ref.~\cite{meng2018}, the level scheme of $^{109}$In has been revised by adding eleven new $\gamma$ rays.

\begin{figure*}
\center
  \includegraphics[width=17cm]{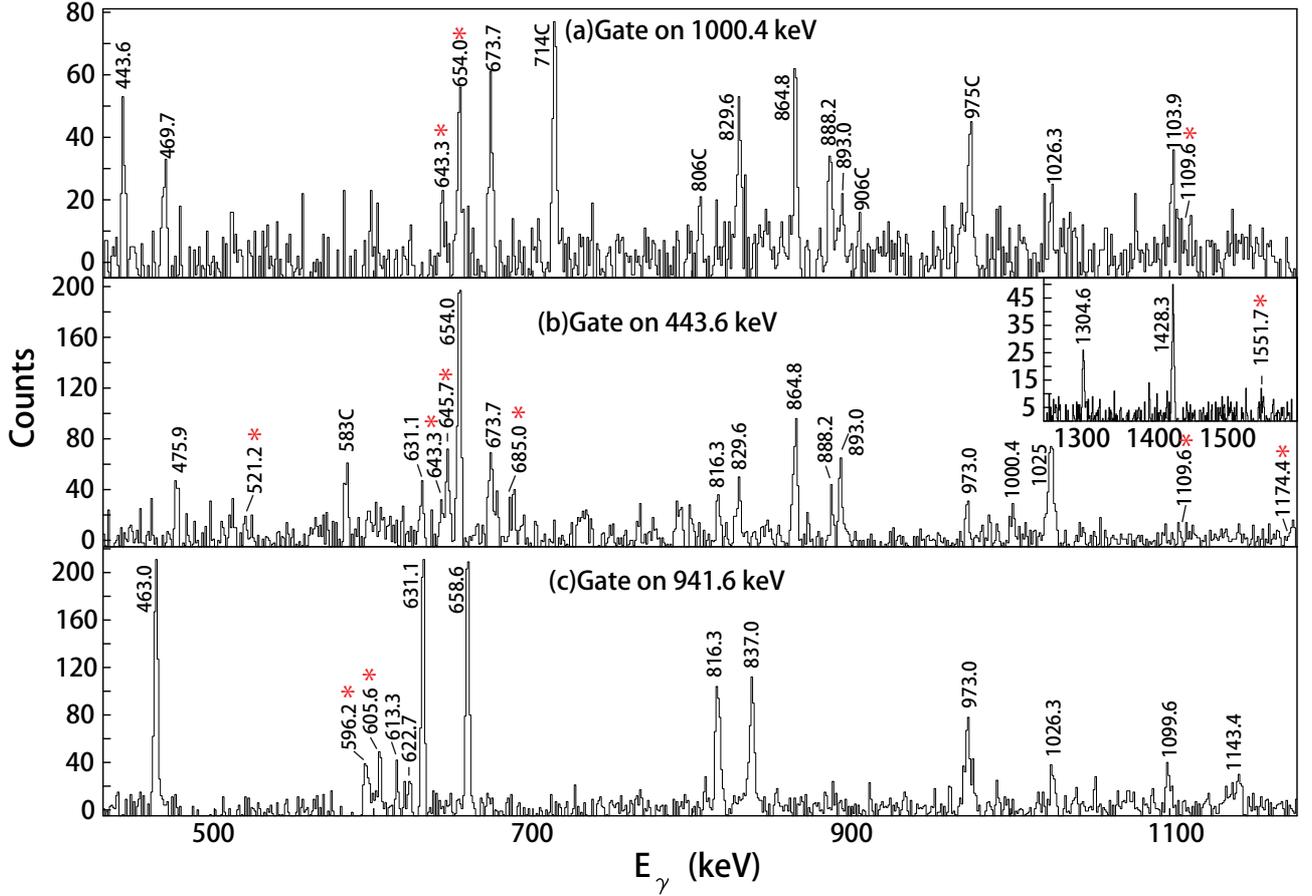}
\caption{(color online) $\gamma$-ray coincidence spectra with gates set on the (a) 1000.4 keV, (b) 443.6 keV, (c) 941.6 transitions. Inset shows the higher-energy part of the spectra (b). The energies marked by the asterisks and C are newly identified $\gamma$ rays and contaminants.}
\label{AMR}
\end{figure*}

Band 7 is an $\Delta I$=2 band and extended to (41/2$^{+}$) state at energy of 8460.8 keV.
In Fig.~\ref{AMR}(a), the 1000.4 keV transition has no coincidence with the newly identified $\gamma$-rays with energies of 1174.4 and 1024.8 keV, while it has coincidene with the 1026.3 keV transition which decays from 11/2$^{+}$ state at 1026.3 keV to 9/2$^{+}$ ground state, and the transitions with energies of 829.6, 888.2, 864.8, 643.3, 1109.6 keV, etc.
In the spectrum gated on 443.6 keV transition, shown in Fig.~\ref{AMR}(b), the transitions decay out from higher levels of band 7 and 9 can be seen, along with the linking transitions with energies of 1109.6, 888.2 keV.
The peak with a centroid energy of 1025 keV can been seen in Fig.~\ref{AMR}(b), which is composed of 1024.8 and 1026.3 keV transitions.
This indicates the existence of 1024.8 keV transition.
Moreover, the 1174.4 and 1024.8 keV transitions have mutual coincidence with the transitions of band 7, but has no coincidence with the 888.2, 1109.6 keV transitions and the $\gamma$ rays in band 9.
Therefore, the 1174.4 and 1024.8 keV transitions are placed on the top of band 7, and the sequence of those transitions is determined by the intensity.
The 864.8 keV transition has no coincidence with the 1109.6 and 643.3 keV transitions.
Such coincident relationship along with the energy summation restricts the position of 643.4 and 1109.6 keV $\gamma$ rays to (31/2$^{+}$) $\rightarrow$ 29/2$^{(+)}$.
The $R_{DCO}$ of 643.4 keV transition is consistent with the $\Delta I$=2 transition, which is similar to the 829.6 and 1000.4 keV transitions of band 9.
Therefore, the 1109.6 keV transition is taken as a linking transition of band 9 and 7, and the 643.4 keV $\gamma$ ray is placed between the state at 7149.8 keV and the state at 6506.5 keV.
The alignment analysis in Sec.~\ref{config} also supports such placement.
In summary, the $\gamma$-rays with the energies of 1024.8, 1174.4 keV belong to band 7.
Band 9 is built on the level at 6506.5 keV and decays to band 7 through the linking transitions with energies of 1109.6, 888.2 keV.

\setlength{\tabcolsep}{14pt}
\begin{table*}
\centering
\caption{The $\gamma$-ray energies, initial- and final-level energies, intensities, DCO ratios, the initial- and final- level spin-parities of $^{109}$In deduced in the present work.}
\label{Table:exp}
\begin{tabular}{ccccccc}
\hline
\hline
$E_{\gamma}$(keV)$^{a}$  &$E_{i}$$\rightarrow$$E_{f}$  &$I_{\gamma}$($\%$)$^{b}$ &$R_{DCO}$(D)$^{c}$ &  $R_{DCO}$(Q)$^{d}$  &$I_{i}^{\pi}$$\rightarrow$$I_{f}$$^{\pi}$&Band\\
\hline
355.7&2447.3$\rightarrow$2091.6&$<$0.1&&&(9/2$^{+}$)$\rightarrow$(5/2$^{+}$)&7 \\
402.0&1428.3$\rightarrow$1026.3 &23(1)&0.61(17)&&13/2$^{+}$$\rightarrow$11/2$^{+}$\\
443.6&4742.8$\rightarrow$4299.2 &4.0(3)&&0.82(17)&25/2$^{(+)}$$\rightarrow$21/2$^{(+)}$&7\\
463.0&5218.7$\rightarrow$4755.7 &2.6(2)&&1.07(16)&27/2$^{+}$$\rightarrow$23/2$^{+}$&8\\
469.7&5396.8$\rightarrow$4927.1 &3(1)&0.64(8)&&29/2$^{(+)}$$\rightarrow$27/2$^{-}$&7$\rightarrow$5\\
475.9&5218.7$\rightarrow$4742.8 &0.7(4)&&0.60(10)&27/2$^{+}$$\rightarrow$25/2$^{(+)}$&8$\rightarrow$7\\
521.2&2968.5$\rightarrow$2447.3 &$<$0.1&&&(13/2$^{+}$)$\rightarrow$(9/2$^{+}$)&7\\
596.2&2318.5$\rightarrow$1722.2 &0.3(2)&&0.9(2)&11/2$^{+}$$\rightarrow$(7/2)$^{+}$&8\\
605.6&2318.5$\rightarrow$1712.5 &0.7(2)&&0.7(2)&11/2$^{+}$$\rightarrow$(9/2)$^{+}$\\
614.2&1712.5$\rightarrow$1099.4&0.2(1)&&&(9/2)$^{+}$$\rightarrow$5/2$^{+}$&\\
623.6&1722.2$\rightarrow$1099.4&0.3(1)&&&(7/2)$^{+}$$\rightarrow$5/2$^{+}$&\\
631.1&5849.8$\rightarrow$5218.7 &2.9(3)&&1.04(11)&31/2$^{+}$$\rightarrow$27/2$^{+}$&8\\
643.3&7149.8$\rightarrow$6506.5 &0.5(2)&&1.0(3)&(35/2$^{+}$)$\rightarrow$(31/2$^{+}$)&9\\
645.7&4299.2$\rightarrow$3653.5&0.8(3)&&1.1(3)&21/2$^{(+)}$$\rightarrow$17/2$^{(+)}$&7\\
654.0&5396.8$\rightarrow$4742.8 &3.2(2)&&0.83(10)&29/2$^{(+)}$$\rightarrow$25/2$^{(+)}$&7\\
658.6&4755.7$\rightarrow$4097.1 &1.2(1)&&0.93(9)&23/2$^{+}$$\rightarrow$19/2$^{+}$&8\\
673.5&2101.8$\rightarrow$1428.3 &69(3)&&&19/2$^{+}$$\rightarrow$13/2$^{+}$\\
673.7&2102.0$\rightarrow$1428.3 &14.7(7)&1.62(23)&&17/2$^{+}$$\rightarrow$13/2$^{+}$\\
685.0&3653.5$\rightarrow$2968.5 &0.3(3)&&1.1(6)&17/2$^{(+)}$$\rightarrow$(13/2$^{+}$)&7\\
816.3&6666.1$\rightarrow$5849.8 &1.5(1)&&1.19(16)&35/2$^{+}$$\rightarrow$31/2$^{+}$&8\\
829.6&7979.4$\rightarrow$7149.8 &1.2(2)&&0.96(29)&(39/2$^{+}$)$\rightarrow$(35/2$^{+}$)&9\\
837.0&3155.5$\rightarrow$2318.5 &1.0(3)&&0.93(7)&15/2$^{+}$$\rightarrow$11/2$^{+}$&8\\
864.8&6261.6$\rightarrow$5396.8 &4.2(3)&&0.80(7)&33/2$^{(+)}$$\rightarrow$29/2$^{(+)}$&7\\
888.2&7149.8$\rightarrow$6261.6 &1.6(2)&&0.78(17)&(35/2$^{+}$)$\rightarrow$33/2$^{(+)}$&9$\rightarrow$7\\
893.0&2995.0$\rightarrow$2102.0 &7.3(7)&&0.46(8)&19/2$\rightarrow$17/2$^{+}$\\
941.6&4097.1$\rightarrow$3155.5 &1.0(1)&&1.10(13)&19/2$^{+}$$\rightarrow$15/2$^{+}$&8\\
973.0&7639.1$\rightarrow$6666.1 &1.6(2)&&0.98(13)&39/2$^{+}$$\rightarrow$35/2$^{+}$&8\\
1000.4&8979.8$\rightarrow$7979.4&0.6(1)&&1.4(5)&(43/2$^{+}$)$\rightarrow$(39/2$^{+}$)&9\\
1024.8&7286.4$\rightarrow$6261.6 &2.6(7)&&&(37/2$^{+}$)$\rightarrow$33/2$^{(+)}$&7\\
1026.3&1026.3$\rightarrow$0 &31(1)&0.59(7)&&11/2$^{+}$$\rightarrow$9/2$^{+}$\\
1099.4&1099.4$\rightarrow$0 &0.5(1)&&&5/2$^{+}$$\rightarrow$9/2$^{+}$&\\
1109.6&6506.5$\rightarrow$5396.8&$<$0.1&&&(31/2$^{+}$)$\rightarrow$29/2$^{(+)}$&9$\rightarrow$7\\
1143.4&8782.5$\rightarrow$7639.1 &0.4(2)&&1.00(26)&43/2$^{+}$$\rightarrow$39/2$^{+}$&8\\
1174.4&8460.8$\rightarrow$7286.4&0.3(2)&&&(41/2$^{+}$)$\rightarrow$(37/2$^{+})$&7\\
1304.2&4299.2$\rightarrow$2995.0 &0.8(1)&&0.54(11)&21/2$^{(+)}$$\rightarrow$19/2\\
1428.3&1428.3$\rightarrow$0 &100&&&13/2$^{+}$$\rightarrow$9/2$^{+}$\\
1551.7&3653.5$\rightarrow$2101.8&0.5(2)&&&17/2$^{(+)}$$\rightarrow$(19/2$^{+})$&\\
\hline
\hline
\end{tabular}
\begin{threeparttable}
\begin{tablenotes}
\item[a)] Uncertainties are between 0.2 and 0.5 keV depending upon their intensity.
\item[b)] Intensities are normalized to the 1428.3 keV transition with $I_{\gamma}=100$.
\item[c)] DCO ratios gated by dipole transitions.
\item[d)] DCO ratios gated by quadrupole transitions.
\end{tablenotes}
\end{threeparttable}
\end{table*}

The DCO of the 888.2 keV transition needs special explanation.
In our early work~\cite{meng2018}, the 888.2 keV transition is thought to be an $E$2 transition and belong to band 7.
Nevertheless, the DCO of the 888.2-keV transition is 0.78(17), which is not a strict proof for an $E$2 transition.
Now the 888.2-keV transition is suggested as a linking transition between band 9 and 7, considering the newly found transitions and coincidence relationships.
The DCO information of the newly found $\gamma$ rays with energy of 1109.6 keV can not been extracted for the weak intensity.
However, if we suppose that the linking transitions with energies of 1109.6 and 888.2 keV are $E$2 transitions, there will be a 37/2$^{+}$ state at energy of 7149.8 keV, which is lower than that of the (37/2$^{+}$) state at 7286.4 keV of band 7. It will be inconsistent with the intensity of the 643.3 and 1024.8 keV transitions.
Therefore, we suggest those linking transitions are $\Delta I$=1 transitions, and the bandhead of band 9 at energy of 6506.5 keV is (31/2$^{+}$).

The newly identified 645.7 and 685.0 keV transitions can been seen in the spectrum gated on 443.6 keV transition, as shown in Fig.~\ref{AMR}(b).
Though the 521.2 and 355.7 keV transitions can not been identified in Fig.~\ref{AMR}(b) for their weak intensities, each of the 645.7, 685.0, 521.2, 355.7 keV transitions has mutual coincidence with their cascade $\gamma$ rays.
Therefore, they are taken as the member of band 7, and the sequence of those four $\gamma$ rays are determined by the intensities.

The 443.6 keV transition has been identified as a $\Delta I$=2 transition in the early work~\cite{meng2018}.
The $R_{DCO}$ of 645.7 and 685.0 keV transition extracted from the spectrum gated on the 443.6 keV transition are around 1, which correspond to $\Delta I$=2 transitions.
While it is difficult to extract the DCO information of the 521.2 and 355.7 keV transitions, we suggest them as $\Delta I$=2 transition considering that they are the intraband transitions of band 7.
The parity of band 7 is suggested to be positive according to the alignment analysis in Sec.~\ref{config}.

Band 8 consists of nine $\Delta I$=2 transitions and is extended to the 43/2$^{+}$ state at 8782.5 keV.
From the 941.6 keV gated spectrum, as shown in Fig.~\ref{AMR}(c), all the members of band 8 can be identified, along with the 605.6, 614.2, 623.6 and 1099.4 keV transitions.
In the early work\\~\cite{meng2018}, the decay path of band 8 is not clear.
With the observed 605.6 and 596.2 keV transitions, the band 8 are connected to the known (9/2)$^{+}$ state at 1712.5 keV and the (7/2)$^{+}$ state at 1722.2 keV~\cite{TOI}.
The 605.6 keV transition is a linking transition, which links the 11/2$^{+}$ state at 2318.5 keV to the (9/2)$^{+}$ state at 1712.5 keV.
Because the band 8 is composed of $\Delta I$=2 transitions, the $\gamma$ ray with energy of 596.2 keV is taken as a member of band 8, and the (7/2)$^{+}$ state at 1722.2 keV is assigned as the bandhead of band 8.
The positive bandhead also supports the parity assignment of band 8 in Ref.~\cite{meng2018}.

\section{DISCUSSION}

\subsection{Systemic discussion and configuration assignment}\label{config}

\begin{figure}
\resizebox{0.5\textwidth}{!}{
\includegraphics{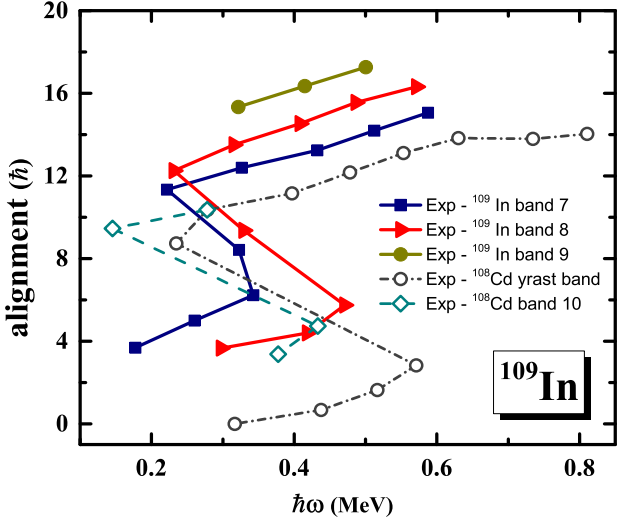}
}
\caption{ (Color online) Experimental alignment as a function of rotational frequency $\hbar$$\omega$ for band 7, 8, 9 in $^{109}$In, together with that of the yrast band and band 10 in $^{108}$Cd, relative to a Harris parametrization of the core of $J_{0}$=7~$\hbar$$^{2}$MeV$^{-1}$ and  $J_{1}$=9~$\hbar$$^{4}$MeV$^{-3}$. }
\label{band7}
\end{figure}

The experimental alignment as a function of rotational frequency for bands 7, 8 and 9 is shown in Fig.~\ref{band7}, and that of the yrast band and band 10 of the neighboring even-even nucleus $^{108}$Cd is shown as a comparison.
The configuration of band 8 is assigned as $\pi g_{7/2}g^{-2}_{9/2}$ before the backbend and $\pi g_{7/2}g^{-2}_{9/2}\otimes\nu h_{11/2}^{2}$ after backbend in Ref.~\cite{meng2018}.
In this work, the bandhead of band 8 has been observed. The whole behaviour of band 8 before the backbend also supports the configuration assignment.

The band 10 of $^{108}$Cd is built on a non-aligned excitation into the $\nu$$h_{11/2}$ subshell~\cite{108Cd}.
Before the backbend, the initial aligned spin of band 7 is nearly 2.5$\hbar$ greater than that of the band 10 of $^{108}$Cd, which can be caused by the occupation of the odd proton in the $d_{5/2}$ orbital.
A sharp backbend in both bands occurs at around 0.28 MeV and with similar gains in aligned spin, which is consistent with $h_{11/2}$ neutron pair alignment.
Therefore, the band 7 before the backbend should build on a non-aligned neutron excitation into the $\nu$$h_{11/2}$ subshell associating with the $\pi$$d_{5/2}$ orbital, and the backbend can be attributed to the alignment of the neutrons in the $h_{11/2}$ orbitals.
Band 9, which decays to band 7 through two linking transitions, can be related to different neutron excitations in comparison with band 7.
The alignments of band 9 is about 3~$\hbar$ greater than that of band 7, which could be attributed to the midshell $g_{7/2}(d_{5/2})$ neutron alignment.

\begin{figure}
\resizebox{0.5\textwidth}{!}{
\includegraphics{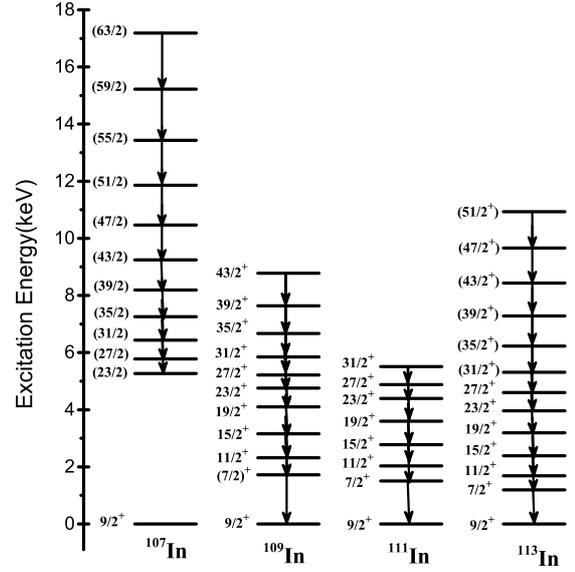}
}
\caption{ Rotational bands involving the $\pi$$g_{7/2}$ orbital in $^{107, 109, 111, 113}$In. Ground states of 9/2$^{+}$ are shown as references.}
\label{energy}
\end{figure}

According to the configuration assignment, bands 7 and 8 are believed to be related to $1p1h$ proton excitations from the $g_{9/2}$ orbital to one of the $g_{7/2}$ and $d_{5/2}$ orbitals above the shell gap.
Similar bands related to the $g_{7/2}$ orbital in $^{107, 111, 113}$In~\cite{SMB_107In,MR_111In,AMR113In} are summarized in Fig~\ref{energy}.
Even though several corresponding levels in $^{107}$In have not been observed, the isotopic regularity of the level energies is significant.
It is worth noting that the excitation energies of $1p1h$ proton excitation from $g_{9/2}$ to $g_{7/2}$ orbital in $^{109, 111, 113}$In are within 1$\sim$2 MeV relative to the ground state, and decreases with the increasing neutron number.
The proton-neutron residual interaction may play an important role in $1p1h$ excitation from $\pi$$g_{9/2}$ to $\pi$$g_{7/2}$ orbital at such low energy.
It reduces the energy spacing between the $\pi$$g_{9/2}$ and $\pi$$g_{7/2}$ orbitals, and its impact is enhanced when more neutrons are occupying the midshell.

\begin{figure}
\resizebox{0.5\textwidth}{!}{
\includegraphics{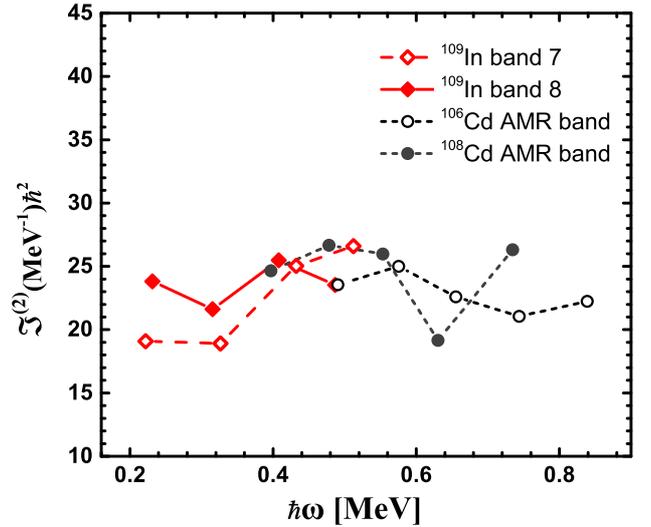}
}
\caption{(Color online) Dynamic moment of inertia $\mathfrak{J}^{(2)}$ as a function of rotational frequency $\hbar$$\omega$ for bands 7 and 8 in $^{109}$In, and the antimagnetic band in $^{106,108}$Cd.}
\label{J2}
\end{figure}

Moreover, when the additional proton of indium nuclei is occupying the $g_{7/2}$ or $d_{5/2}$ orbital, the rotational bands after the $h_{11/2}$ neutrons alignment at high frequencies are perfect candidates for the two-shears-like mechanism, such as the rotational bands in $^{108, 110, 112, 113}$In~\cite{anti108110In,Sun2016,antiMR_112In,AMR113In}.
The dynamic moment of inertia $\mathfrak{J}^{(2)}$ is a sensitive probe of the nuclear collectivity.
The $\mathfrak{J}^{(2)}$ and rotational frequency can be extracted experimentally by the following formulae,

\begin{displaymath}
\hbar\omega_{\rm exp}=\frac{1}{2}E_\gamma(I \rightarrow I-2)
\end{displaymath}
\begin{displaymath}
\mathfrak{J}^{(2)}\approx\frac{dI}{d\omega}=\frac{4}{E_\gamma(I+2 \rightarrow I)-E_\gamma(I \rightarrow I-2)}
\end{displaymath}

$\mathfrak{J}^{(2)}$ of bands 7 and 8 after the backbend in $^{109}$In are shown in Fig.~\ref{J2}.
The typical antimagnetic rotational bands in $^{106}$Cd, $^{108}$Cd~\cite{anti106108Cd,antimr_108Cd} are also shown for comparison.
As shown in Fig.~\ref{J2}, $\mathfrak{J}^{(2)}$ stays around 23~MeV$^ {-1}$$\hbar$$^{2}$ as rotation frequency increases for bands 7 and 8 after backbend, and have the similar pattern with that of AMR bands in $^{106,108}$Cd.
Such small and stable value of $\mathfrak{J}^{(2)}$ indicates that bands 7 and 8 after backbend in $^{109}$In are much less collective, and can be candidates for antimagnetic rotation.

\begin{figure*}[b]
\center
\includegraphics[width=12cm]{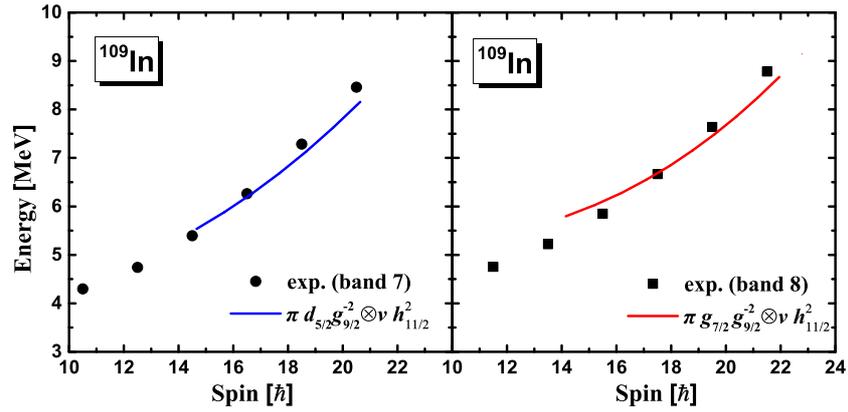}
\caption{(color online) Energy spectrum of bands 7 (left) and 8 (right) after the backbend obtained from the TAC-RMF calculations, in comparison with the corresponding data in~$^{109}$In.}
\label{EI}
\end{figure*}

\begin{figure*}[b]
\center
\includegraphics[width=12cm]{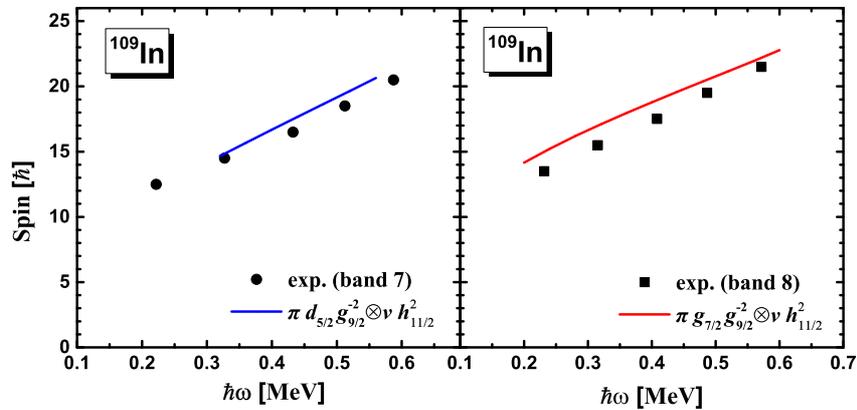}
\caption{(color online) Total angular momentum as a function of the rotational frequency for bands 7 (left) and 8 (right) after the backbend obtained from the TAC-RMF calculations in comparison with the corresponding data in~$^{109}$In.}
\label{spin}
\end{figure*}

\subsection{Theoretical interpretation}

In the following, the rotational structure of two positive parity bands in $^{109}$In are investigated by tilted axis cranking relativistic mean-field (TAC-RMF) approach.
In contrast to its non-relativistic counterparts~\cite{1}, the relativistic mean field (RMF) approach including point-coupling or mesonic exchange interaction~\cite{lj16,lj17,lj18}, takes the fundamental Lorentz symmetry into account from the very beginning so that naturally takes care of the important spin degree of freedom and time-odd fields, resulting in great successes on many nuclear phenomena~\cite{1,2,3,4,5,6}.
Moreover, without any additional parameters, the rotation excitations can be described self-consistently with the tilted axis cranking relativistic mean-field (TAC-RMF) approach~\cite{meng2013progress,6}.
In particular, the TAC-RMF model has been successfully used in describing magnetic rotation (MR) and AMR microscopically and self consistently in different mass regions~\cite{meng2013progress,6}, and
especially the 110 region, such as the AMR bands in $^{105, 109, 110}$Cd~\cite{zhao2012prc,zhao2011prl,zhao2012covariant,peng2015magnetic,zhang2014competition} and $^{108, 110, 112, 113}$In \\ \cite{Sun2016,antiMR_112In,AMR113In}, and also the MR bands in $^{113,114}$In~\cite{113In,MR_114In}.
In the present TAC-RMF calculations, the point-coupling interaction PC-PK1~\cite{zhao2010new} is used for the Lagrangian without any additional parameters.
A basis of 10 major oscillator shells is adopted for the solving of the Dirac equation and pairing correlations are neglected.
In order to describe bands 7 and 8 in $^{109}$In, the configurations  $\pi d_{5/2}g_{9/2}^{-2}\otimes\nu h_{11/2}^2$ and $\pi g_{7/2}g_{9/2}^{-2}\otimes\nu h_{11/2}^2$ are adopted in the TAC-RMF calculations, respectively.

The calculated results for the $\pi d_{5/2}g_{9/2}^{-2}\otimes\nu h_{11/2}^2$ and $\pi g_{7/2}g_{9/2}^{-2}\otimes\nu h_{11/2}^2$ configuration are shown in Fig.~$\ref{EI}$ and Fig.~$\ref{spin}$ in comparison with the experimental data for bands 7 and 8 after the backbend.
It could be seen that the TAC-RMF calculations including both energy spectrum and rotational frequency based on the assigned configurations are in a good agreement with the experimental data, supporting the configuration assignment.

\begin{figure}
\resizebox{0.5\textwidth}{!}{
\includegraphics[scale=0.32]{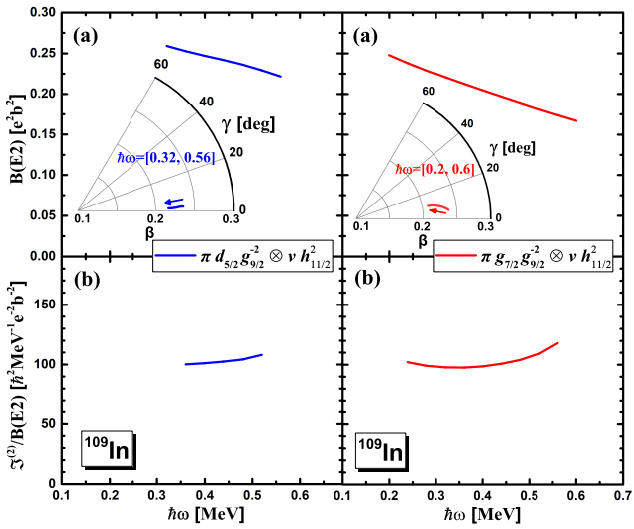}
}
\caption{(color online) $B(E2)$ values~(a) and~$\mathfrak{J}^{(2)}/B(E2)$~ratios (b) as functions of the rotational frequency for bands 7 (left) and 8 (right) in the TAC-RMF calculations for the assigned configurations. Insert: Deformation parameters~$\beta$~and~$\gamma$~driven by the increasing rotational frequency in the TAC-RMF calculations. The arrow indicates the increasing direction of the rotational frequency.}
\label{BJ}
\end{figure}

\begin{figure*}[b]
\center
\includegraphics[width=12cm]{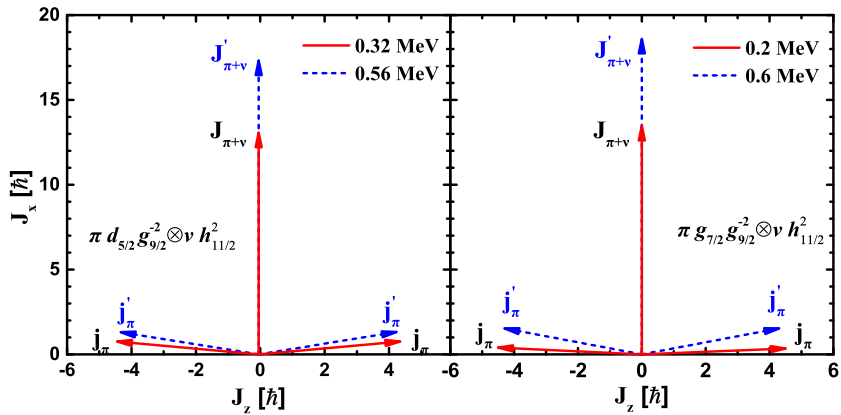}
\caption{(color online) Angular momentum vectors of neutrons and the low-$\Omega$ $d_{5/2}$($g_{7/2}$) proton, $J_{\pi+\nu}$, and the two high-$\Omega$ $g_{9/2}$ proton holes, $j_\pi$, for bands 7 and 8 calculated with TAC-RMF theory.}
\label{twoS}
\end{figure*}

After the backbend in bands 7 and 8, the calculated $\mathfrak{J}^{(2)}$ well reproduce the data and the $\mathfrak{J}^{(2)}$ values are around 20-25~MeV$^ {-1}$$\hbar$$^{2}$, which are much smaller than the typical values ($\sim 35$~MeV$^ {-1}$$\hbar$$^{2}$) for the $A=110$ rigid spherical rotor.
This indicates that bands 7 and 8 are not based on a collective behavior, but most likely an antimagnetic rotation, as discussed in Section.~\ref{config}.

Weak $E2$ transition is one of the typical characteristics of AMR and reflects the small deformation of the core which causes large ratios of $\mathfrak{J}^{(2)}$ to the reduced transition probability $B(E2)$ values.
Furthermore, the $B(E2)$ values decrease with increasing angular momentum rather rapidly.
The $B(E2)$ values and $\mathfrak{J}^{(2)}/B(E2)$ ratios as functions of the rotational frequency in the TAC-RMF calculations for the assigned configurations of bands 7 and 8 are given respectively in Fig.~\ref{BJ}.
The $B(E2)$ values are shown to be decrease smoothly with increasing rotational frequency, while the $\mathfrak{J}^{(2)}/B(E2)$ ratios show rising tendencies for both bands 7 and 8.
It should be noted that the calculated $\mathfrak{J}^{(2)}/B(E2)$ ratios for those two bands are around $100-120\;\hbar^2$MeV$^ {-1}e^{-2}b^{-2}$, which are much higher than that for a typical deformed rotational band ($\sim 10$\\$\;\hbar^2$MeV$^ {-1}e^{-2}b ^{-2}$~\cite{Frauendorf2001}) and also in agreement with the expectations from AMR bands~\cite{zhao2012prc,antiMR_112In,Sun2016}.

The decrease of the $B(E2)$ values can be attributed to the evolution of the nuclear deformation.
As shown in the inset of Fig.~\ref{BJ}(a), with increasing rotational frequency, the nucleus undergoes a smooth decrease in $\beta$ deformation with a rather small and steady triaxiality ($\gamma\leq 10^\circ$) for both bands 7 and 8, which is responsible for the falling tendency of $B(E2)$ values with rotational frequency.

In order to examine the two-shears-like mechanism for bands 7 and 8, $J_{\pi+\nu}$ (the angular momentum
vectors of neutrons and the low-$\Omega$ proton) and $j_\pi$ (the two high-$\Omega$ $g_{9/2}$ proton holes) in the TAC-RMF calculations have been extracted and shown in Fig.~\ref{twoS}.
Taking band 8 with the configuration $\pi g_{7/2}g_{9/2}^{-2}\otimes\nu h_{11/2}^2$ as an example.
The angular momentum $J_{\pi+\nu}$ is related to all the neutron levels and the occupied low-$\Omega$ $g_{7/2}$ proton in the intrinsic system.
At the bandhead ($\hbar\omega=\;$0.2 MeV), the two $j_\pi$ are nearly perpendicular to $J_{\pi+\nu}$ and pointing opposite to each other, which form the blades of the two shears.
As the rotational frequency increases, the gradual alignment of the $g_{9/2}$ proton hole vectors $j_\pi$ toward $J_{\pi+\nu}$ generates angular momentum, while the direction of the total angular momentum stays unchanged.
This leads to the closing of the two shears simultaneously by moving one blade toward the other, demonstrating the two-shears-like mechanism in band 8.
A similar mechanism can also been seen in TAC-RMF calculations with assigned configuration of $\pi d_{5/2}g_{9/2}^{-2}\otimes\nu h_{11/2}^2$ for band 7 as shown in Fig.~\ref{twoS}.

\section{SUMMARY}

In summary, the $\Delta I$=2 rotational bands populated in the $^{100}$Mo($^{14}$N, 5$n$)$^{109}$In reaction have been modified and extended by eleven new $\gamma$ rays.
The systematic discussion has been made and the configurations for the $\Delta I$=2 rotational bands have been assigned.
The dynamic moment of inertia shows that bands 7 and 8 after backbend are much less collective.

The experimental data of bands 7 and 8 in $^{109}$In have been compared with the TAC-RMF calculations, and good agreements have been obtained.
The predicted $B(E2)$, deformation $\beta$ and $\gamma$, as well as $\mathfrak{J}^{(2)}/B(E2)$ ratios in TAC-RMF calculations based on the $\pi d_{5/2}g_{9/2}^{-2}\otimes\nu h_{11/2}^2$ and $\pi g_{7/2}g_{9/2}^{-2}\otimes\nu h_{11/2}^2$ configurations have been discussed and the characteristic features of AMR for the bands 7 and 8 after the backbend have been shown.
The two-shears-like mechanism for bands 7 and 8 show that they can be candidate antimagnetic rotational bands.
Further experimental investigation such as life-time measurements are expected for a conclusive interpretation.

\vspace{24 pt}
We thank the crew of the HI-13 tandem accelerator at the China Institute of Atomic Energy for their help in steady operation of the accelerator and for preparing the target.
This work is partially supported by the National Natural Science Foundation of China under Contracts No. 11375023, No. 11575018, No. U1867210, No. 11675063, U1832211, and 11922501.

%
\bibliographystyle{unsrt}
\bibliography{mybib}

\begin{thebibliography}{10}

\bibitem{Frauendorfcon}
S.~Frauendorf, J.~Meng, and J.~Reif.
\newblock In M.~A. Deleplanque, editor, {\em Proceedings of the Conference on
  Physics from Large $\gamma$-Ray Detector Arrays}, page~52, University of
  California Press, Berkeley, 1994.

\bibitem{CLARK1992247}
R.M. Clark, R.~Wadsworth, E.S. Paul, C.W. Beausang, I.~Ali, A.~Astier, D.M.
  Cullen, P.J. Dagnall, P.~Fallon, M.J. Joyce, M.~Meyer, N.~Redon, P.H. Regan,
  W.~Nazarewicz, and R.~Wyss.
\newblock First observation of a collective dipole rotational band in the
  $\mathrm{A}\sim$ 200 mass region.
\newblock {\em Phys. Lett. B}, 275(3):247 -- 251, 1992.

\bibitem{197Pb}
A.~Kuhnert, M.~A. Stoyer, J.~A. Becker, E.~A. Henry, M.~J. Brinkman, S.~W.
  Yates, T.~F. Wang, J.~A. Cizewski, F.~S. Stephens, M.~A. Deleplanque, R.~M.
  Diamond, A.~O. Macchiavelli, J.~E. Draper, F.~Azaiez, W.~H. Kelly, and
  W.~Korten.
\newblock Oblate collectivity in $^{197}\mathrm{Pb}$.
\newblock {\em Phys. Rev. C}, 46:133--143, Jul 1992.

\bibitem{BALDSIEFEN1992252}
G.~Baldsiefen, H.~Hübel, D.~Mehta, B.V.Thirumala Rao, U.~Birkental,
  G.~Fröhlingsdorf, M.~Neffgen, N.~Nenoff, S.C. Pancholi, N.~Singh,
  W.~Schmitz, K.~Theine, P.~Willsau, H.~Grawe, J.~Heese, H.~Kluge, K.H. Maier,
  M.~Schramm, R.~Schubart, and H.J. Maier.
\newblock Oblate collective bands in $^{199}\mathrm{Pb}$ and
  $^{200}\mathrm{Pb}$.
\newblock {\em Phys. Lett. B}, 275(3):252 -- 258, 1992.

\bibitem{M1_106Ag}
C.~Y. He, L.~H. Zhu, X.~G. Wu, S.~X. Wen, G.~S. Li, Y.~Liu, Z.~M. Wang, X.~Q.
  Li, X.~Z. Cui, H.~B. Sun, R.~G. Ma, and C.~X. Yang.
\newblock Band structures in $^{106}\mathrm{Ag}$ and systematics of shears
  mechanism in the $\mathrm{A}\sim 110$ mass region.
\newblock {\em Phys. Rev. C}, 81:057301, May 2010.

\bibitem{M1_106Ag_cpc}
C.~Y. He, L.~H. Zhu, X.~G. Wu, S.~X. Wen, G.~S. Li, Y.~Liu, Z.~M. Wang, X.~Q.
  Li, X.~Z. Cui, H.~B. Sun, R.~G. Ma, and C.~X. Yang.
\newblock Magnetic rotation in $^{106}\mathrm{Ag}$.
\newblock {\em Chin. Phys. C}, 32(HWL2008-31):120, 2008.

\bibitem{Shapecoexistence106Ag}
Y.~Zheng, L.~H. Zhu, Y.~S. Chen, X.~G. Wu, C.~Y. He, X.~Hao, Y.~Liu, Z.~M.
  Wang, X.~Q. Li, G.~S. Li, Z.~C. Gao, and H.~B. Sun.
\newblock Dramatic transition between electric and magnetic rotations in
  $^{106}\mathrm{Ag}$.
\newblock {\em Sci. China: Phys. Mech. Astron.}, 57(9):1669--1675, Sep 2014.

\bibitem{107Ag}
B.~Zhang, L.~H. Zhu, H.~B. Sun, C.~Y. He, X.~G. Wu, J.~B. Lu, Y.~J. Ma, X.~Hao,
  Yun Zheng, B.~B. Yu, G.~S. Li, S.~H. Yao, L.~L Wang, Chuan Xu, J.~G. Wang,
  and Long Gu.
\newblock New band structures in $^{107}\mathrm{Ag}$.
\newblock {\em Chin. Phys. C}, 35(11):1009, 2011.

\bibitem{M1_107Ag}
S.~H. Yao, H.~L. Ma, L.~H. Zhu, X.~G. Wu, C.~Y. He, Y.~Zheng, B.~Zhang, G.~S.
  Li, C.~B. Li, S.~P. Hu, X.~P. Cao, B.~B. Yu, C.~Xu, and Y.~Y. Cheng.
\newblock Lifetime measurements and magnetic rotation in $^{107}\mathrm{Ag}$.
\newblock {\em Phys. Rev. C}, 89:014327, Jan 2014.

\bibitem{MR_112In_LXQ}
X.~Q. Li, L.~H. Zhu, X.~G. Wu, C.~Y. He, Y.~Liu, B.~Pan, X~Hao, L.~L. H., Z.~M.
  Wang, Li~G. S., Z.~Y. Li, S.~Y. Wang, Q.~Xu, J.~G. Wang, H.~B. Ding, and
  J.~Zhai.
\newblock Shears bands in $^{112}\mathrm{In}$.
\newblock {\em Chin. Phys. C}, 33(HJG2009-67):209, 2009.

\bibitem{MR112In}
C.~Y. He, X.~Q. Li, L.~H. Zhu, X.~G. Wu, Y.~Liu, B.~Pan, X.~Hao, L.~H. Li,
  Z.~M. Wang, G.~S. Li, Z.~Y. Li, S.~Y. Wang, Q.~Xu, J.~G. Wang, H.~B. Ding,
  and J.~Zhai.
\newblock New level scheme and magnetic rotation in $^{112}\mathrm{In}$.
\newblock {\em Nucl. Phys. A}, 834(1):84c -- 86c, 2010.

\bibitem{MR_112In}
C.~Y. He, X.~Q. Li, L.~H. Zhu, X.~G. Wu, B.~Qi, Y.~Liu, B.~Pan, G.~S. Li, L.~H.
  Li, Z.~M. Wang, Z.~Y. Li, S.~Y. Wang, Q.~Xu, J.~G. Wang, H.~B. Ding, and
  J.~Zhai.
\newblock Magnetic rotation in $^{112}\mathrm{In}$.
\newblock {\em Phys. Rev. C}, 83:024309, Feb 2011.

\bibitem{113In}
K.~Y. Ma, J.~B. Lu, D.~Yang, H.~D. Wang, Y.~Z. Liu, J.~Li, L.~H. Zhu, X.~G. Wu,
  Y.~Zheng, and C.~Y. He.
\newblock Level structures of $^{113}\mathrm{In}$.
\newblock {\em Eur. Phys. J. A}, 48(6):82, Jun 2012.

\bibitem{113In_cpl}
K.~Y. Ma, J.~B. Lu, D~Yang, J.~Li, H.~D Wang, Y.~Z. Liu, X.~G. Wu, L.~H. Zhu,
  Y.~Zheng, and C.~Y. He.
\newblock High spin states of $^{113}\mathrm{In}$.
\newblock {\em Chin. Phys. Lett.}, 29(6):062102, 2012.

\bibitem{MR_115In}
Z.~Q. Chen, S.~Y. Wang, L.~Liu, P.~Zhang, H.~Jia, B.~Qi, S.~Wang, D.~P. Sun,
  C.~Liu, Z.~Q. Li, X.~G. Wu, G.~S. Li, C.~Y. He, Y.~Zheng, and L.~H. Zhu.
\newblock High-spin states and possible ``stapler'' band in
  $^{115}\mathrm{In}$.
\newblock {\em Phys. Rev. C}, 91:044303, Apr 2015.

\bibitem{Frauendorf2001}
S.~Frauendorf.
\newblock Spontaneous symmetry breaking in rotating nuclei.
\newblock {\em Rev. Mod. Phys.}, 73:463--514, Jun 2001.

\bibitem{meng2013progress}
J.~Meng, J.~Peng, S.~Q. Zhang, and P.~W. Zhao.
\newblock Progress on tilted axis cranking covariant density functional theory
  for nuclear magnetic and antimagnetic rotation.
\newblock {\em Front. Phys.}, 8:55--79, 2013.

\bibitem{6}
J.~Meng(editor).
\newblock {\em Relativistic Density Functional for Nuclear Structure},
  volume~10.
\newblock World Scientific, Singapore, 2016.
\newblock International Review of Nuclear Physics.

\bibitem{anti105Cd}
Deepika Choudhury, A.~K. Jain, M.~Patial, N.~Gupta, P.~Arumugam, A.~Dhal, R.~K.
  Sinha, L.~Chaturvedi, P.~K. Joshi, T.~Trivedi, R.~Palit, S.~Kumar, R.~Garg,
  S.~Mandal, D.~Negi, G.~Mohanto, S.~Muralithar, R.~P. Singh, N.~Madhavan,
  R.~K. Bhowmik, and S.~C. Pancholi.
\newblock Evidence of antimagnetic rotation in odd-$\mathrm{A}$
  $^{105}\mathrm{Cd}$.
\newblock {\em Phys. Rev. C}, 82:061308, Dec 2010.

\bibitem{anti106Cd}
A.~J. Simons, R.~Wadsworth, D.~G. Jenkins, R.~M. Clark, M.~Cromaz, M.~A.
  Deleplanque, R.~M. Diamond, P.~Fallon, G.~J. Lane, I.~Y. Lee, A.~O.
  Macchiavelli, F.~S. Stephens, C.~E. Svensson, K.~Vetter, D.~Ward, and
  S.~Frauendorf.
\newblock Evidence for a new type of shears mechanism in
  $^{106}\mathrm{C}\mathrm{d}$.
\newblock {\em Phys. Rev. Lett.}, 91:162501, Oct 2003.

\bibitem{anti107Cd}
Deepika Choudhury, A.~K. Jain, G.~Anil Kumar, Suresh Kumar, Sukhjeet Singh,
  P.~Singh, M.~Sainath, T.~Trivedi, J.~Sethi, S.~Saha, S.~K. Jadav, B.~S.
  Naidu, R.~Palit, H.~C. Jain, L.~Chaturvedi, and S.~C. Pancholi.
\newblock Multiple antimagnetic rotation bands in odd-$\mathrm{A}$
  $^{107}\mathrm{Cd}$.
\newblock {\em Phys. Rev. C}, 87:034304, Mar 2013.

\bibitem{anti106108Cd}
A.~J. Simons, R.~Wadsworth, D.~G. Jenkins, R.~M. Clark, M.~Cromaz, M.~A.
  Deleplanque, R.~M. Diamond, P.~Fallon, G.~J. Lane, I.~Y. Lee, A.~O.
  Macchiavelli, F.~S. Stephens, C.~E. Svensson, K.~Vetter, D.~Ward,
  S.~Frauendorf, and Y.~Gu.
\newblock Investigation of antimagnetic rotation in light cadmium nuclei:
  $^{106,108}\mathrm{Cd}$.
\newblock {\em Phys. Rev. C}, 72:024318, Aug 2005.

\bibitem{antimr_108Cd}
P.~Datta, S.~Chattopadhyay, S.~Bhattacharya, T.~K. Ghosh, A.~Goswami, S.~Pal,
  M.~Saha Sarkar, H.~C. Jain, P.~K. Joshi, R.~K. Bhowmik, R.~Kumar,
  N.~Madhavan, S.~Muralithar, P.~V.~Madhusudhana Rao, and R.~P. Singh.
\newblock Observation of antimagnetic rotation in $^{108}\mathrm{Cd}$.
\newblock {\em Phys. Rev. C}, 71:041305, Apr 2005.

\bibitem{anti110Cd}
Santosh Roy, S.~Chattopadhyay, Pradip Datta, S.~Pal, S.~Bhattacharya, R.K.
  Bhowmik, A.~Goswami, H.C. Jain, R.~Kumar, S.~Muralithar, D.~Negi, R.~Palit,
  and R.P. Singh.
\newblock Systematics of antimagnetic rotation in even-$\mathrm{even}$
  $\mathrm{Cd}$ isotopes.
\newblock {\em Phys. Lett. B}, 694(4):322 -- 326, 2011.

\bibitem{anti108110In}
C.~J. Chiara, D.~B. Fossan, V.~P. Janzen, T.~Koike, D.~R. LaFosse, G.~J. Lane,
  S.~M. Mullins, E.~S. Paul, D.~C. Radford, H.~Schnare, J.~M. Sears, J.~F.
  Smith, K.~Starosta, P.~Vaska, R.~Wadsworth, D.~Ward, and S.~Frauendorf.
\newblock Spectroscopy in the $z=49{}^{108,110}\mathrm{In}$ isotopes: Lifetime
  measurements in shears bands.
\newblock {\em Phys. Rev. C}, 64:054314, Oct 2001.

\bibitem{Sun2016}
Wu~Ji Sun, Hai~Dan Xu, Jian Li, Yong~Hao Liu, Ke~Yan Ma, Dong Yang, Jing~Bing
  Lu, and Ying~Jun Ma.
\newblock Antimagnetic rotation in $^{108,110}\mathrm{In}$ with tilted axis
  cranking relativistic mean-field approach.
\newblock {\em Chin. Phys. C}, 40(8):084101, 2016.

\bibitem{antiMR_112In}
X.~W. Li, J.~Li, J.~B. Lu, K.~Y. Ma, Y.~H. Wu, L.~H. Zhu, C.~Y. He, X.~Q. Li,
  Y.~Zheng, G.~S. Li, X.~G. Wu, Y.~J. Ma, and Y.~Z. Liu.
\newblock Candidate antimagnetic rotational band in $^{112}\mathrm{In}$.
\newblock {\em Phys. Rev. C}, 86:057305, Nov 2012.

\bibitem{AMR113In}
K.~Y. Ma, J.~B. Lu, J.~Li, D.~Yang, Y.~J. Ma, W.~J. Sun, X.~Guan, D.~M. Zhang,
  L.~H. Zhu, X.~G. Wu, Y.~Zheng, C.~B. Li, and Y.~Z. Liu.
\newblock Possible antimagnetic rotational band and neutron alignment in
  $^{113}\mathrm{In}$.
\newblock {\em Phys. Rev. C}, 100:014326, Jul 2019.

\bibitem{meng2018}
M.~Wang, Y.~Y. Wang, L.~H. Zhu, B.~H. Sun, G.~L. Zhang, L.~C. He, W.~W. Qu,
  F.~Wang, T.~F. Wang, Y.~Y. Chen, C.~Xiong, J.~Zhang, J.~M. Zhang, Y.~Zheng,
  C.~Y. He, G.~S. Li, J.~L. Wang, X.~G. Wu, S.~H. Yao, C.~B. Li, H.~W. Li,
  S.~P. Hu, and J.~J. Liu.
\newblock New high-spin structure and possible chirality in
  $^{109}\mathrm{In}$.
\newblock {\em Phys. Rev. C}, 98:014304, Jul 2018.

\bibitem{Radford}
D.C Radford.
\newblock $\mathrm{ESCL8R}$ and $\mathrm{LEVIT8R}$: Software for interactive
  graphical analysis of hpge coincidence data sets.
\newblock {\em Nucl. Instrum. Methods Phys. Res. A}, 361(1):297 -- 305, 1995.

\bibitem{TOI}
Richard~B Firestone.
\newblock The table of isotopes-and beyond.
\newblock {\em Transactions of the American Nuclear Society}, 75:117--119,
  1996.

\bibitem{108Cd}
I.~Thorslund, C.~Fahlander, J.~Nyberg, S.~Juutinen, R.~Julin, M.~Piiparinen,
  R.~Wyss, A.~Lampinen, T.~L\"{o}nnroth, D.~M\"{u}ller, S.~T\"{o}rm\"{a}nen,
  and A.~Virtanen.
\newblock The role of the shape driving $\mathrm{h}_{12}$ neutron orbital in
  $^{108}\mathrm{Cd}$.
\newblock {\em Nucl. Phys. A}, 564(2):285 -- 313, 1993.

\bibitem{SMB_107In}
E.~Ideguchi, B.~Cederwall, E.~Ganio\ifmmode~\breve{g}\else \u{g}\fi{}lu,
  B.~Hadinia, K.~Lagergren, T.~B\"ack, A.~Johnson, R.~Wyss, S.~Eeckhaudt,
  T.~Grahn, P.~Greenlees, R.~Julin, S.~Juutinen, H.~Kettunen, M.~Leino, A.~P.
  Leppanen, P.~Nieminen, M.~Nyman, J.~Pakarinen, P.~Rahkila, C.~Scholey,
  J.~Uusitalo, D.~T. Joss, E.~S. Paul, D.~R. Wiseman, R.~Wadsworth, A.~V.
  Afanasjev, and I.~Ragnarsson.
\newblock High-spin intruder band in $^{107}\mathrm{In}$.
\newblock {\em Phys. Rev. C}, 81:034303, Mar 2010.

\bibitem{MR_111In}
P.~Vaska, D.~B. Fossan, D.~R. LaFosse, H.~Schnare, M.~P. Waring, S.~M. Mullins,
  G.~Hackman, D.~Pr\'evost, J.~C. Waddington, V.~P. Janzen, D.~Ward,
  R.~Wadsworth, and E.~S. Paul.
\newblock Particle-hole induced electric and magnetic rotation in
  $^{111}\mathrm{In}$.
\newblock {\em Phys. Rev. C}, 57:1634--1647, Apr 1998.

\bibitem{1}
M.~Bender, P.~H. Heenen, and P.~G. Reinhard.
\newblock Self-consistent mean-field models for nuclear structure.
\newblock {\em Rev. Mod. Phys.}, 75:121--180, Jan 2003.

\bibitem{lj16}
B.~D. Serot and J.~D. Walecka.
\newblock {The Relativistic Nuclear Many Body Problem}.
\newblock {\em Adv. Nucl. Phys.}, 16:1--327, 1986.

\bibitem{lj17}
B.~A. Nikolaus, T.~Hoch, and D.~G. Madland.
\newblock Nuclear ground state properties in a relativistic point coupling
  model.
\newblock {\em Phys. Rev. C}, 46:1757--1781, Nov 1992.

\bibitem{lj18}
W.~H. Long, Giai.~N. V., and J~Meng.
\newblock Density-dependent relativistic hartree–fock approach.
\newblock {\em Phys. Lett. B}, 640(4):150 -- 154, 2006.

\bibitem{2}
P.~Ring.
\newblock Relativistic mean field theory in finite nuclei.
\newblock {\em Prog. Part. Nucl. Phys.}, 37:193 -- 263, 1996.

\bibitem{3}
D.~Vretenar, A.~V. Afanasjev, G.~A. Lalazissis, and P.~Ring.
\newblock Relativistic hartree–bogoliubov theory: static and dynamic aspects
  of exotic nuclear structure.
\newblock {\em Phys. Rep.}, 409(3):101 -- 259, 2005.

\bibitem{4}
J.~Meng, H.~Toki, S.~G. Zhou, S.~Q. Zhang, W.~H. Long, and L.~S. Geng.
\newblock Relativistic continuum hartree bogoliubov theory for ground-state
  properties of exotic nuclei.
\newblock {\em Prog. Part. Nucl. Phys.}, 57(2):470 -- 563, 2006.

\bibitem{5}
J.~Meng and S.~G. Zhou.
\newblock Halos in medium-heavy and heavy nuclei with covariant density
  functional theory in continuum.
\newblock {\em J. Phys. G: Nucl. Part. Phys.}, 42(9):093101, 2015.

\bibitem{zhao2012prc}
P.~W. Zhao, J.~Peng, H.~Z. Liang, P.~Ring, and J.~Meng.
\newblock Covariant density functional theory for antimagnetic rotation.
\newblock {\em Phys. Rev. C}, 85:054310, May 2012.

\bibitem{zhao2011prl}
P.~W. Zhao, J.~Peng, H.~Z. Liang, P.~Ring, and J.~Meng.
\newblock Antimagnetic rotation band in nuclei: A microscopic description.
\newblock {\em Phys. Rev. Lett.}, 107:122501, Sep 2011.

\bibitem{zhao2012covariant}
P.~W. Zhao, J.~Peng, H.~Z. Liang, P.~Ring, J.~Meng, et~al.
\newblock Covariant density functional theory for antimagnetic rotation.
\newblock {\em Phys. Rev. C}, 85(5):054310, 2012.

\bibitem{peng2015magnetic}
J.~Peng and P.~W. Zhao.
\newblock Magnetic and antimagnetic rotation in $^{110}\mathrm{Cd}$ within
  tilted axis cranking relativistic mean-field theory.
\newblock {\em Phys. Rev. C}, 91(4):044329, 2015.

\bibitem{zhang2014competition}
P.~Zhang, B.~Qi, S.~Y. Wang, et~al.
\newblock Competition between antimagnetic and core rotation in
  $^{109}\mathrm{Cd}$ within covariant density functional theory.
\newblock {\em Phys. Rev. C}, 89(4):047302, 2014.

\bibitem{MR_114In}
C.B. Li, J.~Li, X.G. Wu, X.F. Li, Y.~Zheng, C.Y. He, G.S. Li, S.H. Yao, B.B.
  Yu, X.P. Cao, S.P. Hu, J.L. Wang, C.~Xu, and Y.Y. Cheng.
\newblock Magnetic rotation in doubly odd nucleus $^{114}\mathrm{In}$.
\newblock {\em Nucl. Phys. A}, 892(Supplement C):34 -- 42, 2012.

\bibitem{zhao2010new}
P.~W. Zhao, Z.~P. Li, J.~M. Yao, J.~Meng, et~al.
\newblock New parametrization for the nuclear covariant energy density
  functional with a point-coupling interaction.
\newblock {\em Phys. Rev. C}, 82(5):054319, 2010.

\end{thebibliography}


\end{document}